# Parmsurv: a SAS Macro for Flexible Parametric Survival Analysis with Long-Term Predictions

*Han Fu [1], Shahrul Mt-Isa [2], Richard Baumgartner [3], William Malbecq [4]*

[1] College of Public Health, The Ohio State University, Columbus, OH, USA
[2] Biostatistics and Research Decision Sciences (BARDS) Health Technology Assessment (HTA) Statistics, MSD Research Laboratories (MRL), MSD Zurich, Switzerland
[3] BARDS Biometrics Research, MRL, Merck & Co., Inc., Kenilworth, NJ, USA
[4] BARDS HTA Statistics, MRL, MSD Europe INC., Brussels, Belgium

*Abstract*   Health economic evaluations often require predictions of survival rates beyond the follow-up period. Parametric survival models can be more convenient for economic modelling than the Cox model. The generalized gamma (GG) and generalized F (GF) distributions are extensive families that contain almost all commonly used distributions with various hazard shapes and arbitrary complexity. In this study, we present a new SAS® macro for implementing a wide variety of flexible parametric models including the GG and GF distributions and their special cases, as well as the Gompertz distribution. Proper custom distributions are also supported. Different from existing SAS® procedures, this macro not only supports regression on the location parameter but also on ancillary parameters, which greatly increases model flexibility. In addition, the SAS® macro supports weighted regression, stratified regression and robust inference. This study demonstrates with several examples how the SAS® macro can be used for flexible survival modeling and extrapolation.

## 1. Introduction

Health economic evaluations usually require long-term predictions of survival beyond the follow-up period of the trial (e.g. 5 or 10 years), so-called survival extrapolation[1,2]. Although the Cox model is ubiquitous because of its ability to estimate covariate effects regardless of baseline survival, fully parametric survival models imposed with underlying distributional assumptions are more appropriate for survival extrapolation[3]. In addition, parametric models can be more flexible than the Cox model and also apply to the cases where the proportional hazards (PH) assumption is violated. The generalized gamma (GG) and generalized F (GF) distributions are extensive families that contain almost all commonly used distributions in parametric survival models with various hazard shapes and arbitrary complexity[4,5]. However, in the majority of the reported studies, only standard PH or accelerated failure time (AFT) models where only the location parameter of the distributions may depend on covariates have been employed[1,2,6]. In some scenarios where participants in different treatment groups have distinct hazard shapes and/or scales, the survival models should incorporate regression on ancillary parameters in these distributions as well. More flexible models include piecewise models and spline models[7,8], but they usually require specification of the number of cut-offs or knots, which may be subject to debate.

There exists statistical software that can fit the commonly used parametric models for survival analysis. The SAS/STAT® LIFEREG procedure fits the AFT models in the GG family but does not include the GF distribution. It only accommodates AFT regression models for the location parameter, without allowing modeling ancillary parameters. The Stata® *streg* command includes most parametric models (except for the gamma and GF) and allows regression on ancillary parameters but does not support censoring types other than right censoring. Although the *flexsurv* R package[3] supports a variety of flexible parametric models and also allows modeling of ancillary parameters, it does not accommodate stratified regression or robust inference. Additionally, since SAS® is the most commonly used software in the pharmaceutical industry, a SAS® program for estimation and inference of flexible parametric survival models is expected to be helpful for pharmaceutical practitioners.

In this study, we present a new SAS® macro called *paramsurv* for implementing a wide variety of flexible parametric models including the GG and GF distributions and their special cases like log normal and log logistic, as well as the Gompertz model. Proper custom distributions are also supported if any two of the three functions – density, hazard and survival – are provided. Different from existing SAS® procedures, this macro not only accommodates regression on the location parameter, but also on ancillary parameters. In addition, the SAS® macro supports weighted regression, stratified regression and robust inference to mitigate the influence of potential model misspecification. To meet the needs of extrapolation beyond the follow-up period in health technology assessments, this SAS® macro is particularly designed for long-term predictions of survival and hazard rates. If a prediction dataset is supplied by the user, the predicted survival and hazard rates at these covariate values and time points will be displayed in the output, together with graphs of extrapolated survival and hazard curves over time. This article demonstrates with several examples how the SAS® macro can be used for flexible survival modeling and long-term survival prediction. The code for the *paramsurv* SAS® macro has been made publicly available at https://github.com/Merck/Parmsurv.

## 2. Statistical Models

### 2.1 Built-in Distributions

Let $T \sim f(t)$ be a random variable representing lifetime of interest with the cumulative distribution function (CDF) denoted by $F(t)$. The survival distribution of $T$ is given by $S(t) = \mathbb{P}(T > t)$, which describes the probability of still being at risk of the event by time $t$. The hazard function is given by $h(t) = f(t)/S(t)$, representing the instantaneous risk of the event. Widely used statistical distributions in health-economic applications[9] include exponential, Weibull, Gompertz, log-logistic, etc. The GG distribution is an extensive parametric family that includes many commonly used distributions. It has three parameters, including location parameter ($\beta$), scale parameter ($\sigma > 0$) and shape parameter ($\lambda \neq 0$). The location parameter $\beta$ is often the parameter of primary interest. The scale and shape parameters $\sigma$ and $\lambda$ are considered as ancillary parameters, which together determine the shape of the hazard function independent of $\beta$. With different values of $\sigma$ and $\lambda$, the GG family includes four distinct hazard shapes: monotonically increasing and decreasing, as well as bathtub and arc-shaped hazards[4], which has applicability to a wide range of clinical scenarios.

The GF family is an even larger family that includes the GG as well as log logistic. It provides additional flexibility for parametric modeling because of an extra parameter. It has four parameters, including the primary location parameter ($\beta$), and three ancillary parameters – scale parameter ($\sigma > 0$), and $m_1, m_2 > 0$. Prentice[10] proposed an alternative parameterization which produces more stable estimation in model fitting and has been used in standard survival software. This parameterization replaces $m_1$ and $m_2$ with

$$q = \left(\frac{1}{m_1} - \frac{1}{m_2}\right)\left(\frac{1}{m_1} + \frac{1}{m_2}\right)^{-1/2}, \qquad p = \frac{2}{m_1 + m_2}$$

so that $q \in \mathbb{R}$ and $p > 0$. Both parameterizations are included in the macro, with the latter referred to as GENF and the former as GENF_ORIG. Besides the four hazard shapes included in the GG, the GF also supports hazard functions that are decreasing but not monotone[5].

We consider six other parametric distributions besides the GG and the GF, briefly described below. The exponential distribution is the simplest parametric distribution commonly used in survival analysis with only one parameter $\beta$. It is the special case of the GG when $\lambda = \sigma = 1$, with constant hazard over time. The Weibull distribution is the special case of the GG when $\lambda = 1$ and also a generalization of the exponential with two parameters, $\beta$ and $\sigma$. It possesses both AFT and PH properties which makes it a popular choice. When the shape parameter $\lambda$ in the GG approaches zero, the distribution becomes the log-normal distribution where the log of event time follows a normal distribution. The gamma distribution is another special case of the GG when $\sigma = \lambda$. The log-logistic distribution is also commonly used, which is a special case of the GF when $m_1 = m_2 = 1$ (or $q = 0, p = 1$ in the Prentice parameterization). Gompertz is the only distribution we consider that does not belong to the GG or GF family. It is often used to depict the adult human lifespan because its hazard increases in an exponential manner. Some key information of these distributions, including parameters, the survival and density functions, is presented in Table 1.

### 2.2 Adjusting for Covariates

To study the influence of covariates on survival under some distribution, we can connect the covariates to the parameters in this distribution via link functions. Let $\beta$ represent the parameter of primary interest, usually the location parameter, and $\boldsymbol{\eta} = (\eta_1, \dots, \eta_K)$ represent any ancillary parameters in the distribution. The density function can be expressed by $f(t|\beta(\boldsymbol{z}), \boldsymbol{\eta}(\boldsymbol{z}))$ where $\boldsymbol{z}$ is a vector of fixed baseline covariates that are associated with the location parameter and/or the ancillary parameters[3]. Link-transformed linear models $g_0(\beta(\boldsymbol{z})) = b_0^\beta + \boldsymbol{b}^{\beta^T}\boldsymbol{z}$ and $g_k(\eta_k(\boldsymbol{z})) = b_0^{\eta_k} + \boldsymbol{b}^{\eta_k T}\boldsymbol{z}$ for $k = 1, \dots, K$ are considered, where $b_0$ and $\boldsymbol{b}$ are the corresponding intercept and coefficients, respectively. When the parameter is restricted to be positive, such as the scale parameter $\sigma$, the log function is used as the link function $g(\cdot)$ in the macro. Otherwise, the identity function is used. Allowing any parameter to depend on any covariate relaxes the assumption of constant hazard ratio or acceleration factor over time in a PH or AFT model and thus greatly increases the model flexibility.

### 2.3 Likelihood Function

For a fully parametric survival model, the maximum likelihood approach is usually applied to estimate the model parameters. The form of the likelihood function for survival data depends on the censoring status and censoring type of each individual. If subject $i$ has a case weight $w_i$ and is right censored, the individual likelihood for subject $i$ is $L_i(\boldsymbol{\theta}|t_i, w_i, \boldsymbol{z}_i) = S_{\boldsymbol{\theta}}(t_i|\boldsymbol{z}_i)$ for $i = 1, \dots, n$ where $\boldsymbol{\theta}$ contains all parameters in the model. Similarly, if subject $i$ is left censored, the individual likelihood is $1 - S_{\boldsymbol{\theta}}(t_i|\boldsymbol{z}_i)$. In the case of interval censoring between time $t_{1i}$ and $t_{2i}$, the individual likelihood can be

calculated by the difference of the survival probabilities, i.e., $S_\theta(t_{1i}|z_i) - S_\theta(t_{2i}|z_i)$. When the event is observed at $t_i$, the likelihood is just the density function $f_\theta(t_i|z_i)$. The overall log-likelihood function is thus the weighted sum of individual log-likelihoods, given by

$$\log L\left(\boldsymbol{\theta}|\boldsymbol{t}_{\{1,\ldots,n\}}, \boldsymbol{w}_{\{1,\ldots,n\}}, \boldsymbol{z}\right) = \sum_{i=1}^{n} w_i \log L_i\left(\boldsymbol{\theta}|t_i, w_i, \boldsymbol{z}_i\right),$$

where $\boldsymbol{w}_{\{1,\ldots,n\}}$ represents the vector of case weights. When weighting is desired, for example in survey data analysis, the vector of weights can be supplied by the user using the weight argument in the macro that corresponds to a weight column in the dataset. Otherwise, $w_i = 1$ for all subjects by default. The AIC and BIC are given by

$$AIC = -2 \max_{\boldsymbol{\theta}} \log L + 2k,$$

$$BIC = -2 \max_{\boldsymbol{\theta}} \log L + \log(n)\, k,$$

respectively, where $k$ is the number of parameters in the model and $n$ is the sample size.

### 2.4 Statistical Inference

The regular estimator for the covariance matrix of the maximum likelihood estimates (MLE) $\widehat{\boldsymbol{\theta}}$ is given by $\hat{V}(\widehat{\boldsymbol{\theta}}) = \hat{A}^{-1}(\widehat{\boldsymbol{\theta}})$, where $\hat{A}(\widehat{\boldsymbol{\theta}}) = -\frac{\partial^2 \log L(\boldsymbol{\theta})}{\partial \boldsymbol{\theta}^2}$ is the observed information matrix. In the case where an incorrect parametric model is adopted, the "sandwich" estimator has been suggested as a robust variance estimator to mitigate the consequences of model misspecification[11]. The robust sandwich estimator for the covariance matrix is given by

$$\hat{V}(\widehat{\boldsymbol{\theta}}) = \hat{A}^{-1}(\widehat{\boldsymbol{\theta}})\hat{B}(\widehat{\boldsymbol{\theta}})\hat{A}^{-1}(\widehat{\boldsymbol{\theta}}),$$

where $\hat{B}(\widehat{\boldsymbol{\theta}}) = \sum w_i U_i(\widehat{\boldsymbol{\theta}}) U_i'(\widehat{\boldsymbol{\theta}}) w_i$ and $U_i(\boldsymbol{\theta}) = \frac{\partial \log L_i(\boldsymbol{\theta})}{\partial \boldsymbol{\theta}}$ is the contribution from the $i$th observation to the score function. Then the estimated standard errors $\widehat{se}(\widehat{\boldsymbol{\theta}})$ are the square roots of the diagonal elements in the covariance matrix $\hat{V}(\widehat{\boldsymbol{\theta}})$. There is an argument in the macro called robust for the users to specify whether regular or robust standard errors are used for inference. The confidence intervals are given by $\widehat{\boldsymbol{\theta}} \pm z_{1-\alpha/2}\widehat{se}(\widehat{\boldsymbol{\theta}})$, where $z_{1-\alpha/2} = \Phi^{-1}(1 - \alpha/2)$ is the percentile of a standard normal distribution. The value of $\alpha$ can be changed by the user, with a default value of 0.05. T values $t\_val$ are calculated by $\widehat{\boldsymbol{\theta}}/\widehat{se}(\widehat{\boldsymbol{\theta}})$ and p-values are $2 \times \Pr(Z > |t\_val|)$ where $Z$ follows a standard normal distribution.

### 2.5 Stratification

When stratification on certain variable(s) is needed in the survival analysis, estimation and inference will be performed in each stratum $s \in \{1, \ldots, S\}$, separately. Weighted averages can be calculated in order to integrate the stratum-specific estimates for parameters and the corresponding standard errors into a set of overall estimates. For point estimates of parameters, the overall estimate for $\boldsymbol{\theta}$ is $\widehat{\boldsymbol{\theta}} = \sum_{s=1}^{S} \widehat{\omega}_s\, \widehat{\boldsymbol{\theta}}_s$, where $\widehat{\omega}_s$ stands for the weight of stratum $s$ and $\widehat{\boldsymbol{\theta}}_s$ is the parameter estimate in stratum $s$. Correspondingly, the overall estimate for covariance matrix can be approximated by $\hat{V}(\widehat{\boldsymbol{\theta}}) = \sum_{s=1}^{S} \widehat{\omega}_s^2\, \hat{V}_s(\widehat{\boldsymbol{\theta}}_s)$, with $\hat{V}_s(\widehat{\boldsymbol{\theta}}_s)$ being the estimate in stratum $s$[12]. One straightforward weighting approach called sample size weighting is used in this macro, where weights are measured by the proportions of sample size for each stratum, i.e., $\widehat{\omega}_s = n_s/n$. The `strata` argument can be used to supply the column name(s) of the stratification variable(s) in the dataset.

### 3. Computation

### 3.1 Data Format of Time-to-Event Response

Using the macro requires the user to supply a dataset with two columns representing the time-to-event response. Two types of data format are acceptable by the macro for the time-to-event response. The first format includes two columns of time points, $t_1$ and $t_2$, where $t_1$ represents the lower bound of the event interval and $t_2$ represents the upper bound. When only one of them is missing in an observation, the missing value represents either infinity or negative infinity. For example, when an observation in the dataset has a positive value of $t_1$ and a missing $t_2$, the time interval of the event happens becomes $(t_1, \infty)$ and this observation is considered as a right-censored record censored at $t_1$. How different values of $t_1$ and $t_2$ correspond to different censoring types or censoring status is shown in the second column of Table 2. In the cases where both $t_1$ and $t_2$ are missing, or either of them is negative, or $t_1 > t_2$, the observations are considered to be invalid and deleted from the dataset before conducting any analysis.

Since right censoring is most commonly seen in survival analysis, another format is acceptable exclusively for right censoring survival data. It requires two columns in the dataset, observed time $\tilde{T}$ (argument `t1` in the macro) and censoring status $\delta$ (argument `censor`), shown in the third column of Table 2. $\delta_0$ is the value in censoring status $\delta$ that corresponds to censoring, as opposed to observed events. By default, $\delta_0 = 0$ corresponds to censoring observations. The user can specify $\delta_0$ in the `censval` argument. When $\tilde{T} < 0$ or $\delta$ is missing, the observation will be deleted. In practice, the users need to specify the column name for either `censor` or `t2`, which determines which format will be used to load the data. Inside the program, the time-to-event responses in both formats will be converted to the first format in `t1` and `t2` for consistency.

### 3.2 Initialization

If initial values are not supplied by the user, parameters in the optimization will be automatically initialized. Simple summary statistics of the event or censoring times are used for the initial values of $\beta$ and $\sigma$ in the distributions within the GG and GF families. We denote the observed time by $\tau$ (given in the fourth column in Table 2 except for interval censored data where we approximate $\tau$ by $(t_1 + t_2)/2$). The mean and standard deviation of $\{\log \tau_1, \ldots, \log \tau_n\}$ are then used as the initial values of $\beta$ and $\sigma$, respectively. Parameters that are restricted to be non-zero, such as $\lambda$ in the GG and $q$ in GF, are initialized as 0.5. For other parameters that are restricted to be positive, the initial values are set to be 1. In a custom distribution, the mean value of $\{\log \tau_1, \ldots, \log \tau_n\}$ is used to initialize the location parameter $\beta$, 1 used to initialize any positive parameters, and 0 used for others. Despite the availability of automatic initialization, we recommend the users to supply proper initial values using the `init` argument if any prior information is known.

### 3.3 Log Transformation

Parameters constrained to be positive will go through log transformations during the optimization to yield more stable convergence results. For example, the scale parameter $\sigma$ is constrained to be positive in most of the built-in distributions. Instead of directly optimizing $\sigma$, we let $\log \sigma$ which is unconstrained enter the optimization and transform the optimization results back to the original scale afterwards. The estimated standard errors for these parameters of the original scale are calculated using the Delta's method. The confidence intervals are calculated by taking the exponential of the intervals in the log scale. An argument called `log_result` determines whether the estimation results in the log scale will be printed in the output.

### 3.4 Implementation

The SAS/IML® matrix programming language was used for computation. The non-linear programming (NLP) tool in SAS/IML® has been used for maximum likelihood estimation, where the optimization algorithm can be specified by the user, including Newton-Raphson (default), quasi-Newton, and trust region. The users can specify the amount of output printed by the NLP subroutine using the `nlp_print` argument (integer value from 0 to 5) to check the optimization process. A higher value corresponds to more detailed output. By default, no output is generated for the NLP process (`nlp_print=0`). To calculate the score function and the information matrix required in variance estimation, a subroutine called NLPFDD in SAS/IML® was used to obtain approximate derivatives by finite differences.

## 4. Example Applications

In this section, we present several examples to demonstrate how the macro can be used for flexible survival modeling and long-term survival prediction. A dataset is generated with a right-censoring survival response and two covariates, age and sex, for 100 individuals. Age ($x_1$) is a continuous covariate that is drawn from a normal distribution with mean of 50 and standard error of 10, and has been centered and scaled. Sex ($x_2$) is a character covariate drawn from a Bernoulli distribution with the probability of 0.5, with 1 representing female and 0 representing male. The event time $T_i$ follows a $GG(\beta_i, \sigma, \lambda)$ distribution, where $\beta_i = b_1 x_{1i} + b_2 x_{2i}$ with $b_1 = -1$ and $b_2 = 0.5$. The scale parameter $\sigma$ and shape parameter $\lambda$ are both 1 which essentially corresponds to an exponential distribution. The censoring time $C_i$ follows an independent exponential distribution with the mean equaling to the average event time of all subjects. The observed time $t_i$ for subject $i$ is then the smaller value of $T_i$ and $C_i$, and the censoring status $\delta_i$ is an indicator of whether the event was observed ($T_i \leq C_i$). Assume this dataset is stored as `data` in the current workspace, whose first 10 observations are shown in Table 3. The `time` column represents the observed time $t_i$ and the `delta` column represents the censoring status $\delta_i$ where 0 is censored and 1 is observed. Several examples are presented below to call the macro with this dataset.

### 4.1 Built-in Model Fitting and Model Comparison

Various built-in models with different distributions can be fitted to this example dataset. To fit such models, the users need to specify the dataset name (`data=data`) and variable names that correspond to the observed time (`t1=time`), censoring status (`censor=delta`), and covariates (`covars=age sex`). Then any built-in distribution can be applied. The macro calls to fit the exponential and Weibull models are shown as below and those for other distributions are similar. Only the location parameters are expected to depend on the covariates age and sex in these models. The parameter estimates, inference statistics, and the values of log-likelihood, AIC and BIC, will be printed in the output after these macro calls, shown in Table 4 and 5, respectively.

```
%paramsurv(data=data, t1=time, censor=delta, covars=age sex, dist=exp)
```

```
%paramsurv(data=data, t1=time, censor=delta, covars=age sex, dist=Weibull)
```

Table 6 summarizes the log-likelihood values, AIC and BIC for different models for model comparison. The exponential model has the lowest BIC because it is the closest to the underlying true model, and the Weibull model achieves the lowest AIC which has weaker penalization on model complexity than BIC.

Note in the models above, male is automatically selected as the reference group. The users can change the reference group(s) for classification covariate(s) by specifying the `refgrp` argument when calling the macro, for example, `refgrp=%str(female)`. A non-numeric covariate in the dataset will be automatically considered as a classification covariate, like `sex` here. If the users want to specify numeric variables as classification covariates, they can achieve that using the `class_cov` argument, for example, `class_cov=x1 x2`, where covariates are delimited by a space. When there are multiple classification covariates and the users want to specify the reference groups, they need to specify them for all the covariates, using the syntax like `refgrp = %str(placebo,high risk)` where covariates are delimited by a comma and the reference level for a single covariate may contain spaces. The order in `refgrp` should be consistent as they appear in `covars` or `class_cov`. When there are $K$ levels in a classification covariate with $K > 2$, $(K-1)$ dummy variables will be created in a small built-in macro called `%class_cov_prep`, where a new numeric covariate matrix is prepared for the subsequent analyses.

### 4.2 Regression on Ancillary Parameters

More complex models can be fitted where not only the location parameter but also ancillary parameters depend on certain covariates. For example, the users may want to fit a GG model where the scale parameter $\sigma$ is associated with age and sex and the shape parameter $\lambda$ is associated with sex. If we denote age by $x_1$ and sex by $x_2$, the density can be expressed by $f(t \mid \beta(x_1, x_2), \sigma(x_1, x_2), \lambda(x_2))$. Since $\sigma$ in the GG is constraint to be positive, the covariates are linked to the parameter $\sigma$ via the log link function. For $\beta$ and $\lambda$, the identity link function is used. The relationships between the covariates and the parameters are expressed by: $\beta(x_1, x_2) = \gamma_0 + \eta_{01}x_1 + \eta_{02}x_2$, $\log(\sigma(x_1, x_2)) = \gamma_1 + \eta_{11}x_1 + \eta_{12}x_2$, and $\lambda(x_2) = \gamma_2 + \eta_{22}x_2$. This model can be fitted by specifying the `anc` argument to be `anc=%str(sigma(age sex),lambda(sex))`, as in the following macro call. Different parameters are delimited by commas and the covariate names are expected inside the parentheses that follow the corresponding parameters.

```
%paramsurv(data=data, t1=time, censor=delta, covars=age sex,
    anc=%str(sigma(age sex), lambda(sex)), dist=gengamma)
```

The output of this model is given in Table 7. We can now obtain the estimated coefficients that are associated with different parameters. In this complex model, only the age effect on the location parameter $\beta$ is highly significant with a very small p-value, which is somewhat consistent with the underlying true model. With the model complexity increased, the AIC and BIC are higher than those in the previous results, as expected.

### 4.3 Custom Distribution

In this example, we illustrate how to fit a model with a user-defined distribution which is essentially a Weibull model with the PH parameterization. In this parameterization, the hazard and survival functions are expressed by $h(t) = \mu\alpha t^{\alpha-1}$ and $S(t) = \exp(-\mu t^\alpha)$, respectively, where $\mu = \exp(-\beta)$ is a function of the location parameter $\beta$ and $\alpha$ is the new scale parameter. The hazard thus satisfies the PH property with the baseline hazard being $\alpha t^{\alpha-1}$. To define a custom distribution, any two of the three functions – density, hazard and survival – need to be provided (or, alternatively, the log of those functions if computation can be simplified in this way). The third function will be automatically calculated inside the program according to the known two functions. In this example, we can fit the model by specifying the custom hazard and log survival functions as `hazard=%str(mu*alpha*time**(alpha-1))` and `log_survival=%str(-mu*time**alpha)`. Note that the time-to-event variable is denoted by `time` in the expressions.

The users need to let the macro know which parameters are ancillary parameters by specifying `param_anc=alpha`. Since $\alpha$ is constraint to be positive, the users might want to transform $\alpha$ to the log scale during optimization by specifying `log_transf_param=alpha`. As an alternative, the users can specify a lower bound of zero for the parameter $\varphi$ by providing the `lower` argument and thus avoid log-transformation. (There is also an `upper` argument similarly.) But the log-transformation is recommended for a positive parameter for better convergence. We still use `beta` to be the location parameter which is the default option; it can be specified via the `location` argument otherwise. The location parameter $\beta$ is automatically linked to the covariates age and sex via an identity function in this case. The users need to define the relationship between the undefined parameter in the distribution, $\mu$, and the built-in defined parameter $\beta$ by providing the argument of `custom_prep=%str(mu=exp(-beta);)`. This relationship is then used in a small built-in macro called `%param_prep` to connect the parameters in the optimization to the parameters or coefficients in the model.

```
%paramsurv(data=data, t1=time, censor=delta, covars= age sex,
   hazard=%str(mu*alpha*time**(alpha-1)), log_survival=%str(-
   mu*time**alpha), param_anc=alpha, log_transf_param=alpha,
   custom_prep=%str(mu=exp(-beta);))
```

Table 8 displays the output of this user-defined model. It has the same log-likelihood value, AIC and BIC as the Weibull model in Table 5, because they are essentially the same model with different parameterizations.

### 4.4 Long-Term Prediction

This example demonstrates how to perform long-term survival prediction based on a particular model. A GG model is fitted to the dataset, based on which we would like to make predictions for cases with certain covariate values. To mitigate potential influence of model-misspecification, the users can specify `robust=yes` to use the robust estimator to estimate the standard errors. For the purpose of illustration, we use a simple prediction dataset called `pred` stored in the current workspace, shown in Table 9. It needs to contain a column named `time` that contains the time points at which the users want to predict. It should also contain all covariates used in the model, if there are any, either in the `covars` or `anc` argument, and the column names for the covariates should be exactly the same as the original dataset. Each row is a prediction case. The prediction dataset will go through the same procedure of preparing a numerical data matrix in the small macro `%class_cov_prep` as the original dataset. The survival and hazard functions are defined in the `surv` and `haz` functions to return the predicted values of survival and hazard, respectively, using the `%param_prep` macro as well. Again, the NLPFDD subroutine is used to calculate the first derivatives which are used to estimate the predicted standard errors for survival and hazard using the Delta's method. In this toy example, the prediction dataset indicates that we would like to predict the survival and hazard values at time point 1 for males and females at average age (scaled age = 0), respectively.

For the predicted survival and hazard trajectory plots, we need to remove the rows in the prediction dataset that have duplicated covariate values, so that each row is a unique prediction case that corresponds to a line in the plots. The users can specify the longest prediction time $t_L$ for the curves, which is set to 5 here (`pred_max_time=5`). One hundred ticks from $t_L/100$ to $t_L$ are used to generate the plots. For each of the 100 time points, the predicted survival and hazard as well as the confidence intervals are calculated and restored. The `pred_plot_cl` argument indicates whether point-wise confidence bands are shown in the plots (`yes` by default). If there are multiple unique prediction cases in the prediction dataset or there is stratification, the group labels for each line in the plots will be generated.

```
%paramsurv(data=data, t1=time, censor=delta, covars=age sex,
    dist=gengamma, robust=t, pred=pred, pred_max_time=5)
```

The fitting results are shown in Table 10 where an additional table for prediction is presented. We can see that females at the average age are predicted to have higher survival probabilities and lower hazard, which is consistent with the data generating process. The extrapolation plots for survival and hazard are shown in Figure 1 where the two prediction cases form two groups, males and females at the average age. In this example, females at average age are predicted to have better survival rates than males at all times, while the confidence bands of predicted hazard for females are too wide to claim a significant difference compared with the hazard for males.

## 5. Conclusion

In this article, we introduced a SAS® macro called `%paramsurv` for general parametric survival analysis that allows various parametric distributions of arbitrary complexity. The macro accommodates fixed continuous or classification covariates that are associated with primary or ancillary parameters, which is not available in the existing SAS® procedures. The present macro further allows case weights, stratification and robust inference, and most importantly, supports long-term predictions of survival and hazard rates, which is essential for health economic evaluations. This macro provides appropriate tools to facilitate the effective applications of flexible parametric survival modelling, especially in the regulated drug development environment.

**Disclaimers**

H. Fu was employed as an intern at Merck Sharp & Dohme Corp., a subsidiary of Merck & Co., Inc., Kenilworth, N.J., U.S.A during the study conduct. S. Mt-Isa, R. Baumgartner, and W. Malbecq are employees of Merck Sharp & Dohme Corp., a subsidiary of Merck & Co., Inc., Kenilworth, N.J., U.S.A.

**Acknowledgement**

The authors acknowledge the contribution of Mr. Harris Kampouris, who is an employee of Merck Sharp & Dohme Corp., a subsidiary of Merck & Co., Inc., Kenilworth, N.J., U.S.A., in providing oversight and support to the programming activities throughout the study.

Table 1. Survival and density functions of built-in distributions

| Distribution | Survival Function $S(t)$ | Density Function $f(t)$ |
|---|---|---|
| `Exp` / `Exponential` ($\beta$) | $e^{-\gamma t}$ where $\gamma = e^{-\beta}$ | $\gamma e^{-\gamma t}$ where $\gamma = e^{-\beta}$ |
| `Weibull` ($\beta, \sigma > 0$) | $\exp[-(\gamma t)^{1/\sigma}]$ where $\gamma = e^{-\beta}$ | $\frac{1}{\sigma t}(\gamma t)^{1/\sigma}\exp[-(\gamma t)^{1/\sigma}]$ where $\gamma = e^{-\beta}$ |
| `Gamma` ($\beta, \sigma > 0$) | $1 - cdf('gamma', t, \sigma^{-2}, \sigma^2 e^{\beta})$ | $pdf('gamma', t, \sigma^{-2}, \sigma^2 e^{\beta})$ |
| `Lnorm` (Log-Normal, $\beta, \sigma > 0$) | $1 - cdf('logn', t, \beta, \sigma)$ | $pdf('logn', t, \beta, \sigma)$ |
| `Gompertz` ($\beta, \nu \ne 0$) | $\exp[-\gamma(e^{\nu t} - 1)/\nu]$ where $\gamma = e^{-\beta}$ | $\gamma e^{\nu t}\exp[-\gamma(e^{\nu t} - 1)/\nu]$ where $\gamma = e^{-\beta}$ |
| `Llogis` (Log-logistic, $\beta, \sigma > 0$) | $\dfrac{1}{1 + e^{-\beta\sqrt{2}/\sigma}t^{\sqrt{2}/\sigma}}$ | $\dfrac{\sqrt{2}e^{-\beta\sqrt{2}/\sigma}t^{\sqrt{2}/\sigma}}{t\sigma\left[1 + (e^{-\beta}t)^{\sqrt{2}/\sigma}\right]^2}$ |
| `GenGamma` ($\beta, \sigma > 0$, $\lambda \ne 0$) | $\begin{cases} 1 - cdf\left('gamma', t^{\frac{\lambda}{\sigma}}, \lambda^{-2}, \lambda^2 e^{\frac{\beta\lambda}{\sigma}}\right), if\ \lambda > 0 \\ cdf\left('gamma', t^{\frac{\lambda}{\sigma}}, \lambda^{-2}, \lambda^2 e^{\frac{\beta\lambda}{\sigma}}\right), if\ \lambda < 0 \end{cases}$ | $\dfrac{\|\lambda\|}{\sigma t}t^{\frac{\lambda}{\sigma}}\,pdf\left('gamma', t^{\frac{\lambda}{\sigma}}, \lambda^{-2}, \lambda^2 e^{\frac{\beta\lambda}{\sigma}}\right)$ |
| `GenF_orig` ($\beta, \sigma > 0$, $m_1 > 0$, $m_2 > 0$) or `GenF` ($\beta, \sigma > 0$, $q, p > 0$) | $cdf\left('beta', \dfrac{m_2}{m_2 + m_1 e^w}, m_2, m_1\right)$ where $e^w = (e^{-\beta}t)^{\delta/\sigma}$, $\delta = \sqrt{q^2 + 2p}$, $m_1 = 2/[q^2 + 2p + q\delta]$, $m_2 = 2/[q^2 + 2p - q\delta]$ | $\dfrac{\delta e^{-\frac{\beta m_1 \delta}{\sigma}} t^{\left(\frac{\delta}{\sigma}\right)m_1}\left(\frac{m_1}{m_2}\right)^{m_1}}{t\sigma B(m_1, m_2)\left[1 + \left(\frac{m_1}{m_2}\right)(e^{-\beta}t)^{\frac{\delta}{\sigma}}\right]^{(m_1+m_2)}}$ where $\delta = \sqrt{q^2 + 2p}$, $m_1 = 2/[q^2 + 2p + q\delta]$, $m_2 = 2/[q^2 + 2p - q\delta]$ |

$cdf$ and $pdf$ functions in the table correspond to the CDF and PDF functions in SAS®.
$'gamma'$, $'logn'$ and $'beta'$ correspond to gamma, log-normal and beta distributions, respectively.
$B(x, y) = \int_0^1 t^{x-1}(1-t)^{y-1}dt$ is the beta function.

Table 2. Acceptable data format of the time-to-event response inputted by users

| Data Type | Format 1 | Format 2 | Observed Time $\tau$ |
|---|---|---|---|
| Right censored | $t_1 > 0, t_2 = .$ | $\tilde{T} > 0, \delta = \delta_0$ | $t_1$ or $\tilde{T}$ |
| Left censored | $t_1 = ., t_2 > 0$ | N/A | $t_2$ |
| Interval censored | $0 < t_1 < t_2$ | N/A | $(t_1, t_2)$ |
| Event observed | $t_1 = t_2 > 0$ | $\tilde{T} > 0, \delta \neq \delta_0$ | $t_1 = t_2$ or $\tilde{T}$ |

Table 3. First 10 observations in the example dataset

| Obs | time | delta | age | sex |
|---|---|---|---|---|
| 1 | 0.50007 | 1 | -1.03832 | female |
| 2 | 2.92097 | 0 | 0.45638 | female |
| 3 | 2.94312 | 0 | 1.25355 | male |
| 4 | 0.21046 | 1 | -2.13443 | female |
| 5 | 0.05548 | 1 | 0.55603 | female |
| 6 | 2.92410 | 0 | 0.65567 | female |
| 7 | 0.43712 | 1 | -0.44044 | male |
| 8 | 0.44384 | 0 | -0.34079 | male |
| 9 | 0.82329 | 0 | -0.44044 | male |
| 10 | 0.27194 | 1 | -0.73938 | female |

Table 4. Output for the exponential model $f(t|\beta(x_1, x_2))$

| Parameter Estimates | | | | | | |
|---|---|---|---|---|---|---|
| Variable | Estimate | S.E. | 95% CI Lower | 95% CI Upper | t Value | Pr>|t| |
| Intercept | 0.37456 | 0.20213 | -0.02161 | 0.77073 | 1.85 | 0.0639 |
| age | -1.24983 | 0.13301 | -1.51053 | -0.98913 | -9.40 | <.0001 |
| sex_female | 0.44936 | 0.27424 | -0.08815 | 0.98687 | 1.64 | 0.1013 |

| Log-likelihood | AIC | BIC |
|---|---|---|
| -57.656 | 121.312 | 129.128 |

Table 5. Output for the Weibull model $f(t|\beta(x_1, x_2), \sigma)$

**Parameter Estimates**

| Variable | Estimate | S.E. | 95% CI Lower | 95% CI Upper | t Value | Pr>|t| |
|---|---|---|---|---|---|---|
| Intercept | 0.34344 | 0.16734 | 0.01545 | 0.67142 | 2.05 | 0.0401 |
| age | -1.19596 | 0.11362 | -1.41865 | -0.97327 | -10.53 | <.0001 |
| sex_female | 0.44717 | 0.22638 | 0.00347 | 0.89087 | 1.98 | 0.0482 |
| SIGMA | 0.82480 | 0.08442 | 0.67488 | 1.00802 | 9.77 | <.0001 |

| Log-likelihood | AIC | BIC |
|---|---|---|
| -56.043 | 120.086 | 130.507 |

Table 6. Model comparison of built-in distributions

| Distribution | Log-likelihood | AIC | BIC |
|---|---|---|---|
| Exponential | -57.656 | 121.312 | 129.128 |
| Weibull | -56.043 | 120.086 | 130.507 |
| Gamma | -56.204 | 120.408 | 130.829 |
| Log-Normal | -60.688 | 129.375 | 139.796 |
| Gompertz | -57.249 | 122.499 | 132.920 |
| Log-logistic | -58.903 | 125.806 | 136.227 |
| Generalized Gamma | -55.998 | 121.996 | 135.022 |
| Generalized F | -55.998 | 123.997 | 139.628 |

Table 7. Output for the generalized Gamma model $f(t \mid \beta(x_1, x_2), \sigma(x_1, x_2), \lambda(x_2))$

**Parameter Estimates**

| Variable | Estimate | S.E. | 95% CI Lower | 95% CI Upper | t Value | Pr>|t| |
|---|---|---|---|---|---|---|
| beta: intercept | 0.41382 | 0.24427 | -0.06493 | 0.89257 | 1.69 | 0.0902 |
| beta: age | -1.16948 | 0.10990 | -1.38488 | -0.95408 | -10.64 | <.0001 |
| beta: sex_female | 0.33720 | 0.38377 | -0.41497 | 1.08937 | 0.88 | 0.3796 |
| sigma: intercept | -0.31780 | 0.23973 | -0.78767 | 0.15207 | -1.33 | 0.1850 |
| sigma: age | -0.00279 | 0.09426 | -0.18753 | 0.18196 | -0.03 | 0.9764 |
| sigma: sex_female | 0.21777 | 0.35858 | -0.48503 | 0.92057 | 0.61 | 0.5436 |
| lambda: intercept | 1.20726 | 0.55042 | 0.12846 | 2.28605 | 2.19 | 0.0283 |
| lambda: sex_female | -0.30016 | 0.80753 | -1.88290 | 1.28258 | -0.37 | 0.7101 |

| Log-likelihood | AIC | BIC |
|---|---|---|
| -55.798 | 127.596 | 148.438 |

Table 8. Output for the model with custom distribution $f(t \mid \beta(x_1, x_2), \alpha)$

| Parameter Estimates | | | | | | |
|---|---|---|---|---|---|---|
| Variable | Estimate | S.E. | 95% CI Lower | 95% CI Upper | t Value | Pr>\|t\| |
| Intercept | 0.41639 | 0.20511 | 0.01438 | 0.81840 | 2.03 | 0.0423 |
| age | -1.45000 | 0.17901 | -1.80085 | -1.09915 | -8.10 | <.0001 |
| sex_female | 0.54215 | 0.27980 | -0.00624 | 1.09055 | 1.94 | 0.0527 |
| ALPHA | 1.21242 | 0.12409 | 0.99205 | 1.48174 | 9.77 | <.0001 |

| Log-likelihood | AIC | BIC |
|---|---|---|
| -56.043 | 120.086 | 130.507 |

Table 9. The prediction dataset for the example in Section 4.4

| Obs | sex | time | age |
|---|---|---|---|
| 1 | female | 1 | 0 |
| 2 | male | 1 | 0 |

Table 10. Output tables for the prediction example in Section 4.4

| Parameter Estimates | | | | | | |
|---|---|---|---|---|---|---|
| Variable | Estimate | S.E. | 95% CI Lower | 95% CI Upper | t Value | Pr>\|t\| |
| Intercept | 0.38686 | 0.22066 | -0.04563 | 0.81934 | 1.75 | 0.0796 |
| age | -1.19087 | 0.11106 | -1.40854 | -0.97319 | -10.72 | <.0001 |
| sex_female | 0.45092 | 0.22116 | 0.01746 | 0.88438 | 2.04 | 0.0415 |
| SIGMA | 0.78948 | 0.14416 | 0.55196 | 1.12920 | 5.48 | <.0001 |
| LAMBDA | 1.12076 | 0.41671 | 0.30402 | 1.93749 | 2.69 | 0.0072 |

| Log-likelihood | AIC | BIC |
|---|---|---|
| -55.998 | 121.996 | 135.022 |

| Prediction | | | | | | |
|---|---|---|---|---|---|---|
| Obs | sex | time | age | Survival | Survival SE | Hazard | Hazard SE |
| 1 | female | 1 | 0 | 0.687 | 0.058 | 0.449 | 0.104 |
| 2 | male | 1 | 0 | 0.523 | 0.072 | 0.789 | 0.175 |

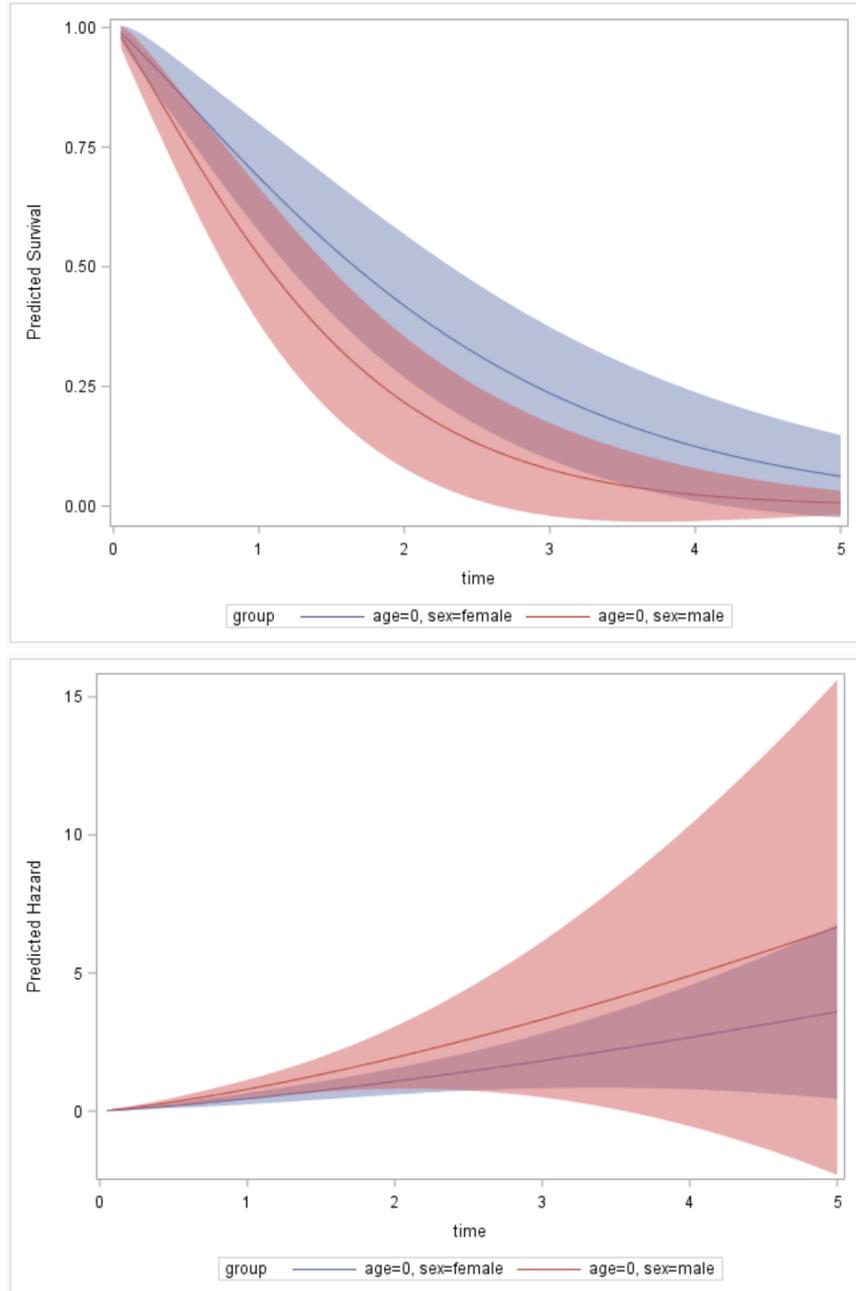

Figure 1. Predicted survival (top) and hazard (bottom) trajectory plots for males and females at average age (scaled age = 0) based on a generalized Gamma model.